\documentclass{jnmp03c}

\begin{document}
\setcounter{page}{139}

\renewcommand{\evenhead}{P G L Leach, S Moyo, S Cotsakis and R L Lemmer}
\renewcommand{\oddhead}{Symmetry, Singularities
and Integrability in Complex Dynamics III}

\thispagestyle{empty}

\FistPageHead{1}{\pageref{leach-firstpage}--\pageref{leach-lastpage}}{Article}

\copyrightnote{2001}{P G L Leach, S Moyo, S Cotsakis and R L Lemmer}

\Name{Symmetry, Singularities and Integrability\\
in Complex Dynamics III:\\
Approximate Symmetries and Invariants}\label{leach-firstpage}

\Author{P G L LEACH~$^{\dag^1 \dag^2}$,
S MOYO~$^{\dag^2}$, S COTSAKIS~$^{\dag^1\dag^2}$ and
R L LEMMER~$^{\dag^1 \dag^3}$}

\Adress{$^{\dag^1}$Laboratory for Geometry, Dynamical Systems and Cosmology
 (GEODYSYC)\\
~~Department of Mathematics, University of the Aegean,
Karlovassi 83 200, Greece\\[2mm]
$^{\dag^2}$School of Mathematical and Statistical Sciences, University of Natal\\
~~Durban 4041, Republic of South Africa\\[2mm]
$^{\dag^3}$Citadel Investment Services, Pretoria, Republic of South Africa}

\Date{Received July 1, 2000; Revised August 17, 2000; Accepted August 21, 2000}

\renewcommand{\theequation}{\thesection.\arabic{equation}}

\setcounter{equation}{0}

\begin{abstract}
\noindent The different natures of approximate symmetries and
their corresponding first integrals/invariants are delineated in
the contexts of both Lie symmetries of ordinary differential
equations and Noether symmetries of the Action Integral.
Particular note is taken of the effect of taking higher orders of
the perturbation parameter. Approximate symmetries of approximate
first integrals/invariants and the problems of calculating them
using the Lie method are considered.
\end{abstract}

\section{Introduction}
According to Gazizov \cite{gazizov}
the first publication devoted to approximate symmetries was by Baikov
et. al.~\cite{baikov1} (see also~\cite{baikov2}).  The concept of an approximate
symmetry goes back to at least the late seventies~\cite{johnson}
 when the observation was made that, since the differential
equations which arise in mathematical modelling are invariably approximate,
one should in fact be considering approximate symmetries.  The observation may
have been a consequence of a number of papers on applications in Cosmology in
the sixties and seventies (cf.~\cite{Komar1,Komar2,Matzner1,Spero1,Spero2}).
 These applications continue~\cite{Zalaletdinov}.
More recently in the context of differential equations the concept
was regarded as sufficiently important for Head~\cite{head} to add
a procedure for the calculation of approximate symmetries to his
Program LIE in~1997.  The procedure is for computations to the
first order in the perturbation parameter~$\epsilon$, but the
program is easily adapted
 for the computation to be made to
any order (subject to the capacity of the machine).

We recall that a differential equation has a Lie symmetry if there exists an
infinitesimal transformation which leaves the differential equation unchanged.
The infinitesimal transformation
\be
\ba{l}
\bar{x} =  x + \alpha \xi,
\vspace{1mm}\\
\bar{y} =  y + \alpha\eta,
\ea \label{1a}
\ee
where $\alpha$ is the infinitesimal parameter, $x$ is the independent variable
and $y$ the dependent variable (the extension to more than one of each is
simple enough in principle), can be written as
\be
\bar{x} = (1 + \alpha G)x, \qquad \bar{y} = (1 + \alpha G)y, \label{2a}
\ee
where $G$ is the differential operator
\be
G = \xi\p_x + \eta\p_y. \label{3a}
\ee
To deal with differential equations the operator $G$ must be extended to give
the infinitesimal transformations in the derivatives induced by the
infinitesimal transformations in the independent and dependent variables.  The
$n$th extension of $G$, which deals with all derivatives up to the $n$th
derivative, is~\cite{leach1}
\be
G^{[n]} = G + \sum_{j=1}^{n}\eta^{[j]}\p_{y^{(j)}}, \label{4a}
\ee
where
\be
\ba{l}
\ds \eta^{[k]} =  \frac{{\rm d}}{{\rm d}x} \eta^{[k-1]} - y^{(k)}\frac{{\rm d}\xi}{{\rm d}x},
\qquad k=1,n,
\vspace{3mm}\\
\ds  \eta^{[0]} = \eta. \ea                    \label{5a} \ee An
$n$th order differential equation, \be
E\left(x,y,y',\ldots,y^{(n)}\right) = 0, \label{6a} \ee is
invariant under the infinitesimal transformation generated by $G$
if \be G^{[n]}E_{\ds {|_{E=0}}}=0. \label{7a} \ee Each
differential operator $G$, with this property is called a symmetry
of the differential equation and the aggregation of all such
symmetries forms the Lie algebra of the differential equation.  A
first integral/invariant, $I(x,y,y',\ldots,y^{(n-1)})$, is
associated with a symmetry~$G$~if \be
G^{[n-1]}I=0\qquad\mbox{and}\qquad \frac{{\rm d}I}{{\rm d}x}=0.
\label{8a} \ee
Instead of a differential equation one can consider the Action Integral as the
object to maintain its invariance under an infinitesimal transformation.  In
this case we have Noether's Theorem~\cite{noether} which states that $G$ is a
Noether symmetry of a Lagrangian, $L\left(x,y,y',\ldots,y^{(n)}\right)$, if there exists
a function~$f$, such that
\be
f' = G^{[n]} L + \xi' L. \label{9a}
\ee
The existence of the symmetry is independent of the Euler--Lagrange equation.
When the Variational Principle is invoked to give the Euler--Lagrange equations,
(\ref{9a}) can be manipulated to give the first integral/invariant
\be
I = f - \left[ \xi L + \left(\eta_{i} - y'_{i}\xi\right) \right] \frac{\p L}{\p y'_{i}}.
\label{10a}
\ee
All of these considerations have been made without reference to
the nature of the coefficient functions $\xi$ and $\eta$.  In
Lie's original work on the symmetries of differential equations he
considered both point~\cite{liedgl} and contact
transformations~\cite{lieber} while Noether used generalized
transformations {\it ab initio}.  There is in fact {\it a priori}
no restriction on the functional dependence of the coefficient
functions and increasingly use is made of generalized and nonlocal
symmetries to expand the classes of problems which can be solved.
If algebraic properties are the primary consideration, it is
better to stay with point or contact symmetries, otherwise there
is no need for constraint.  In the end the choice of type of
infinitesimal transformation to be used is determined by a
utilitarianism appropriate to the occasion~\cite{leach2}.

No matter the type of infinitesimal transformation under
consideration, it is always possible to introduce the approximate
form of such a transformation if the differential equation under
investigation contains a small constant so that the equation can
be regarded as a perturbation of the equation with the constant
set at zero.  Suppose that the differential equation is \be
E\left(x,y,y',\ldots,y^{(n)},\epsilon\right) = 0, \label{11a} \ee
where $\epsilon$ is the small constant and should not be confused
with the $\alpha$ used to denote the infinitesimal parameter in
(\ref{1a}) and (\ref{2a}).  The generator of an approximate
infinitesimal transformation of order $k$ can be written as \be G
= \sum_{i=0}^{k}\epsilon^iG_i, \label{12a} \ee where \be \frac{\p
G_i}{\p \epsilon} = 0.  \label{13a} \ee The generator (\ref{12a})
will be an approximate symmetry of (\ref{11a}) if \be
G^{[n]}E_{{\ds |_{E=0}}}= O\left(\epsilon^{k+1}\right).
\label{14a} \ee We note that, when $n=0$ in (\ref{11a}), we have
the case of {\it a function} and the definition of an invariant
associated with a symmetry group is generalized to that of an
approximate invariant or, more correctly, an invariant function of
an approximate symmetry group.  (In the language of jet bundles,
cf.~Cotsakis, Leach \& Pantazi \cite{Hara}, one is extending the
``approximate'' vector field from one on the manifold --- in the
case of a function --- to that on the $n^{th}$ jet bundle of the
total space.  A differential equation is then a variety living in
the $n^{th}$ jet bundle and in our case we may reinterpret this as
the unperturbed entity of the $\epsilon$-perturbed family of
varieties related to the approximate symmetries.)

Noether's Theorem can be adjusted as follows.
We consider transformations generated by symmetries of the form (\ref{12a}) so
that (\ref{2a}) becomes, to order $k$ in $\epsilon$,
\be
\bar{x} = x + \alpha\sum_{i=0}^{k}\epsilon^i\xi_i,\qquad \bar{y} = y +
\alpha\sum_{i=0}^{k}\epsilon^i\eta_i, \label{nn2}
\ee
where
\begin{equation}
G_i = \xi_i\p_x+\eta_i\p_y,\qquad \frac{\p \xi_i}{\p \epsilon} = 0 =
\frac{\p \eta_i}{\p \epsilon},\qquad
i=0,k.
\label{nn3}
\end{equation}
From Noether's theorem we have that (\ref{12a}) is an approximate
Noether symmetry of the action integral
\be
A = \int_{x_{0}}^{x_{1}} L(x,y,y',\alpha) \, {\rm d} x \label{nn4}
\ee
if it leaves the action integral (\ref{nn4}) unchanged, i.e.
\be
\int_{\bar{x}_{0}}^{\bar{x}_{1}} L\left(\bar{x},\bar{y},\bar{y}',\alpha\right){\rm d} x =
 \int_{x_{0}}^{x_{1}} L\left(x,y,y',\alpha\right){\rm d} x
+ \alpha F +O\left(\epsilon^{k+1}\right), \label{nn5}
\ee
where $\alpha F$ is the infinitesimal contribution from the boundary terms and,
naturally, is written in terms of an expansion in $\epsilon$ as
\begin{equation}
\alpha F = \alpha\sum_{i=0}^{k}\epsilon^i F_i,
\qquad \frac{\p F_i}{\p \epsilon} = 0 = \frac{\p F_i}{\p \alpha}.
\label{nn4a}
\end{equation}
The use of (\ref{nn2}) in (\ref{nn5}) up to order $\epsilon^k$ leads to
the Killing-type equation \cite{sc81}
\be
\ba{l}
\ds \sum_{i=0}^{k}\epsilon^iF'_i = \sum_{i=0}^k
\epsilon^i\left\{\xi_i\frac{\p L_0}{\p x}
+\eta_i\frac{\p L_0}{\p y}+
\left(\eta'_i-y'\xi'_i\right) \frac{\p L_0}{\p y'} +\xi'_iL_0\right\}
\vspace{3mm}\\
\ds \phantom{ \sum_{i=0}^{k}\epsilon^iF'_i =}+
\sum_{i=1}^k\epsilon^i\left\{\xi_i\frac{\p L_1}{\p x}+\eta_i\frac{\p L_1}{\p y}+
\left(\eta'_i-y'\xi'_i\right) \frac{\p L_1}{\p y'} +\xi'_iL_1\right\} + O\left(\epsilon^{k+1}\right),
\ea\label{nn6}\hspace{-10mm}
\ee
where prime denotes total
differentiation with respect to $x$, we have written $L=L_0+\epsilon L_1$ in an
obvious notation and all terms of order greater than $\epsilon^k$
are neglected.
Hence (\ref{nn3}) is an approximate  Noether symmetry of (\ref{nn4}) if
the Killing type equation (\ref{nn6}) is satisfied.

Nothing has been said about the dependence of the functions $\xi_i$,
$\eta_i$ and $f_i$ apart from their lack of dependence upon $\epsilon$.
If one seeks approximate point
transformations, these functions depend on $x$ and $y$ only. Otherwise the
functions can be allowed to depend on $x$, $y$ and derivatives of $y$.  Indeed
there is no reason why one should not contemplate nonlocal approximate
symmetries.
A theory of approximate Noether symmetries based on the concept of approximate
infinitesimal transformations can be found in~\cite{knn98}.  However, that
theory is limited to the first order in~$\epsilon$ only and does not lead to the
results presented in this paper.

The rest of this paper, which is the third in a
series~\cite{Flessas,II} devoted to the investigation of the
connections between the three main topics in dynamics, {\it viz}
symmetry, singularities and integrability, topics which
superficially are unrelated in their mathematics and yet are
intimately intertwined, is organized as follows.  In the next
section we examine three simple second order ordinary differential
equations for approximate symmetries.  The equa\-tions~are
\begin{itemize}
\topsep0mm
\partopsep0mm
\parsep0mm
\itemsep0mm
\item[(i)] the simple harmonic oscillator treated as a perturbation of the free particle,
\item[(ii)] the Ermakov--Pinney equation treated as a perturbation of the equation for the
simple harmonic oscillator and
\item[(iii)] an autonomous Emden--Fowler
 equation of order two considered as a perturbation
of the free particle equation.
\end{itemize}

Each of these equations illustrates certain aspects of the properties of
approximate symmetries.  We have chosen simple equations so that the properties
of the symmetries will not be obscured by complexity of the equations.  In \S~3
we consider the properties of approximate Noether symmetries applied to the
Lagrangians of the differential equations above.  In \S~4 we examine some of
the problems which arise in the determination of approximate first
integrals/invariants using the Lie method and the approximate symmetries
associated with them and in \S~5 we present our
observations and conclusions.

\setcounter{equation}{0}

\section{Approximate Lie symmetries: Three case studies}
\subsection{The simple harmonic oscillator}
We write the differential equation as
\be
\ddot{x}+\epsilon x = 0, \label{1}
\ee
where $\epsilon$ is a small parameter.  Naturally this is somewhat artificial since
any value can be used as the coefficient of $x$ by the simple expedient of
rescaling time.  We recall that the number of exact symmetries is eight no
matter the value of $\epsilon$.  The calculation of the symmetries is
straightforward and we
list the approximate symmetries at $O(\epsilon)$, $O\left(\epsilon^2\right)$ and
$O\left(\epsilon^3\right)$.  They are
\begin{eqnarray}
\hspace*{-9mm}
\ba{ll}
O(\epsilon):& \vspace{2mm}\\
G_1= x\p_x, &
G_2 = \epsilon\left(2 t \p_t + x\p_x\right),
\vspace{2mm}\\
\ds G_3 = \left(1-\frac 12 \epsilon t^2\right)\p_x, &
\ds G_4 = \left(t-\frac{1}{3!}\epsilon t^3\right)\p_x,
\vspace{2mm}\\
\ds G_5 = x\left(t-\frac{1}{3!}\epsilon t^3\right)\p_t + x^2\left(1-\frac{1}{2!}
\epsilon t^2\right)\p_x,&
\ds G_{6} = \left(t^2-\frac{1}{3}\epsilon t^4\right)\p_t + x\left(t-\frac{2}{3}t^3\epsilon
\right)\p_x,
\vspace{2mm}\\
\ds G_7 = x\left(1-\frac 12\epsilon t^2\right)\p_t -  tx^2\epsilon\p_x,&
G_8 = \epsilon x\p_t,
\vspace{2mm}\\
G_9 = \left(1-2\epsilon t^2\right)\p_t - 2tx\epsilon\p_x, &
\ds G_{10} = \left(t-\frac{2}{3}\epsilon t^3\right)\p_t + x\left(1-t^2\right)\p_x,
\vspace{2mm}\\
G_{11} = \epsilon \p_t, & G_{12} = \epsilon\p_x,
\vspace{2mm}\\
G_{13} = \epsilon t\p_t, & G_{14} = \epsilon t\p_x,
\vspace{2mm}\\
G_{15} = \epsilon\left(tx\p_t + x^2\p_x\right),
& G_{16} = \epsilon\left(t^2\p_t + tx\p_x\right);
\ea\hspace{-10mm}\nonumber  \\       \label{122}
\end{eqnarray}
\[\hspace*{-9mm}
\ba{ll}
O\left(\epsilon^2\right):\mbox{\raisebox{0mm}[15pt][0pt]{}}&
\vspace{2mm}\\
G_1 = x\p_x, & G_{2} = \epsilon^2 \left(2 t\p_t + x\p_x\right),
\vspace{2mm}\\
\ds G_3 = \left(1-\frac 12\epsilon t^2+\frac{1}{4!}\epsilon^2t^4\right)\p_x,
\hspace{2.2cm}  &
\ds G_4 = \left(t-\frac{1}{3!}\epsilon t^3+\frac{1}{5!}\epsilon^2 t^5\right)\p_x,
\ea
\]
\begin{eqnarray}\hspace*{-9mm}
\ba{ll}
\multicolumn{2}{l}{\ds G_5 = \epsilon\left[ x\left(t-\frac{1}{3!}\epsilon t^3\right)\p_t
+ x^2\left(1-\frac{1}{2!}\epsilon t^2\right)\p_x\right],}
\vspace{2mm}\\
\multicolumn{2}{l}{\ds G_{6} = \left(t^2-\frac{1}{3}\epsilon t^4+\frac{2}{45}\epsilon^2 t^6\right)\p_t +
x\left(t-\frac{2}{3}\epsilon t^3+\frac{2}{15}\epsilon^2t^5\right)\p_x,}
\vspace{2mm}\\
\multicolumn{2}{l}{\ds G_7 = x\left(1-\frac 12 \epsilon t^2+ \frac{1}{4!}\epsilon^2t^4\right)\p_t
- x^2\left(\epsilon t-\frac{1}{3!}\epsilon^2t^3\right)\p_x,}
\vspace{2mm}\\
\ds
G_8 = \epsilon\left[ x\left(1-\frac 12\epsilon t^2\right)\p_t - \epsilon tx^2\p_x\right],
&
\vspace{2mm}\\
\multicolumn{2}{l}{\ds G_9 = \left(1-2\epsilon t^2+\frac{2}{3}\epsilon^2t^4\right)\p_t
 - x\left(2\epsilon t-\frac{4}{3}\epsilon^2t^3\right)\p_x,}
\vspace{2mm}\\
\ds G_{10} = \epsilon\left[ \left(t-\frac{2}{3}\epsilon t^3\right)\p_t + x\left(1-t^2\right)\p_x
\right],\qquad  & G_{11} = \epsilon^2\p_t,
\vspace{2mm}\\
\ds G_{12} = \epsilon\left(1-\frac 12 \epsilon t^2\right)\p_x, &
\ds G_{13} = \left(\epsilon t-\frac{2}{3}\epsilon^2 t^3\right)\p_t - \epsilon^2 t^2 x\p_x,
\vspace{2mm}\\
\ds G_{14} = \epsilon\left(t - \frac{1}{3!}\epsilon t^2\right)\p_x, &
G_{15} = \epsilon^2\left(tx\p_t+x^2\p_x\right),
\vspace{2mm}\\
\ds  G_{16} = \epsilon^2\left(t^2\p_t+tx\p_x\right), &
\vspace{2mm}\\
\multicolumn{2}{l}{\ds G_{17} = \epsilon\left(\frac{2}{3}t^3
-\frac{2}{15}\epsilon t^5\right)\p_t
+ \epsilon x\left(t^2-\frac{1}{3}\epsilon t^4\right)\p_x,}
\vspace{2mm}\\
\multicolumn{2}{l}{\ds G_{18} = x\left(t-\frac{1}{3!}\epsilon t^3+\frac{1}{5!}\epsilon^2t^5\right)\p_t
+ x^2\left(1-\frac 12 \epsilon t^2+
\frac{1}{4!}\epsilon^2 t^4\right)\p_x,}
\vspace{2mm}\\
G_{19} = \epsilon^2\p_x, &
G_{20} = \epsilon^2 x\p_t +\epsilon x\p_x,
\vspace{2mm}\\
G_{21} = \epsilon^2 t\p_t, & G_{22} = \epsilon^2t\p_x,
\vspace{2mm}\\
\ds G_{23} = \epsilon\left[\left(1-2\epsilon t^2\right)\p_t - 2\epsilon tx\p_x\right], &
\ds G_{24} = \epsilon\left[\left(1-2\epsilon t^2\right)\p_t - 2\epsilon tx\p_x\right];
\ea \hspace{-10mm}              \nonumber\\        \label{133}
\end{eqnarray}
\[\hspace*{-9mm}
\ba{ll}
O\left(\epsilon^3\right):\mbox{\raisebox{0mm}[15pt][0pt]{}}&
\vspace{2mm}\\
G_1 = x\p_x,&
G_{2} = \epsilon^3\left(2t \p_t + x\p_x\right),
\vspace{2mm}\\
\ds G_{3} = \left(1 - \frac 12 \epsilon t^2 + \frac{1}{4!} \epsilon^2 t^4 - \frac{1}{6!}
 \epsilon^3 t^6\right)\p_x,\hspace{11mm} &
\ds G_{4} = \left( t - \frac{1}{3!} \epsilon t^3 + \frac{1}{5!} \epsilon^2 t^5
- \frac{1}{7!} \epsilon^3 t^7\right)\p_x,
\vspace{2mm}\\
\multicolumn{2}{l}{\ds G_{5} = \epsilon^2 \left[x\left(t - \frac{1}{3!} \epsilon t^3\right)\p_t + x^2
\left(1  - \frac{1}{2!} \epsilon t^2\right)\p_x \right],}
\vspace{2mm}\\
\multicolumn{2}{l}{\ds G_{6} = \epsilon \left(t^2 - \frac{1}{3} \epsilon t^4
+ \frac{2}{45} \epsilon^2 t^6\right)\p_t+ \epsilon x \left( t
- \frac{2}{3}  \epsilon t^3 + \frac{2}{15} \epsilon^2 t^5\right)\p_x,}
\vspace{2mm}\\
\multicolumn{2}{l}{\ds G_{7} = x\left( 1 - \frac{1}{2!} \epsilon t^2 + \frac{1}{4!}
\epsilon^2 t^4 - \frac{1}{6!} \epsilon^3
t^6\right)\p_t- x^2\left(\epsilon t - \frac{1}{3!} \epsilon^2 t^3 + \frac{1}{5!} \epsilon^3
t^5\right)\p_x,}
\vspace{2mm}\\
\ds G_{8} = \epsilon^2 \left[ x \left( 1 - \frac 12 \epsilon t^2\right)\p_t
-x^2 \epsilon t\p_x \right], &
\ea\hspace{-10mm}\nonumber
\]
\begin{eqnarray}\hspace*{-9mm}
\ba{ll}
\multicolumn{2}{l}{\ds G_{9} = \left(1 - 2 \epsilon t^2 + \frac{2}{3} \epsilon^2 t^4
- \frac{4}{45} \epsilon^3 t^6\right)\p_t
-\left(2 x \epsilon t - \frac{4}{3} x \epsilon^2 t^3 + \frac{4}{15} x \epsilon^3 t^5\right)\p_x,}
\vspace{2mm}\\
\multicolumn{2}{l}{\ds G_{10} = \left( t^2 - \frac{1}{3} \epsilon t^4 + \frac{2}{45}
\epsilon^2 t^6 - \frac{1}{315} \epsilon^3 t^8\right)\p_t
+\left(x^2 - \frac{1}{2!} x^2 \epsilon t^2 + \frac{1}{4!} x^2 \epsilon^2 t^4 -
\frac{1}{6!} x^2 \epsilon^3 t^6\right)\p_x,}
\vspace{2mm}\\
\ds G_{11} = \epsilon^3\p_t, &
\ds G_{12} = \epsilon \left(1 - \frac 12 \epsilon t^2 + \frac{1}{4!} \epsilon^2 t^4\right)\p_x,
\vspace{2mm}\\
\ds G_{13} = \left(2 \epsilon^2 t - \frac{4}{3} \epsilon^3 t^3\right)\p_t
- 2 x \epsilon^3 t^2 \p_x, \qquad &
\ds G_{14} = \epsilon^2\left(t-\frac{1}{3!}\epsilon t^3\right)\p_x,
\vspace{2mm}\\
\ds G_{15} = \epsilon^3\left(tx\p_t + x^2\p_x\right),
&\ds G_{16} = \epsilon^3\left(t^2\p_t + tx\p_x\right),
\vspace{2mm}\\
\multicolumn{2}{l}{\ds G_{17} = \left( x t - \frac{1}{3!} x \epsilon t^3 + \frac{1}{5!}
x \epsilon^2 t^5 - \frac{1}{7!} x \epsilon^3 t^7\right)\p_t
+ \left( x t - \frac{2}{3} x \epsilon t^3 + \frac{2}{15} x \epsilon^2 t^5 - \frac{4}{315}
x \epsilon^3 t^7\right)\p_x,}
\vspace{2mm}\\
\multicolumn{2}{l}{\ds G_{18} = \left( 2 \epsilon t - \frac{4}{3} \epsilon^2 t^3
+ \frac{4}{15} \epsilon^3 t^5\right)\p_t
+\left(\epsilon x + \frac{2}{3} x \epsilon^3 t^4 - 2 x t^2 \epsilon^2\right)\p_x,}
\vspace{2mm}\\
\ds G_{19} = \epsilon^3\p_x, &
\ds G_{20} = \epsilon^2\left(1- \frac 12 \epsilon t^2\right)\p_x,
\vspace{2mm}\\
\ds G_{21} = \epsilon^3 t\p_t, &
\ds G_{22} = \epsilon^3 t\p_x,
\vspace{2mm}\\
\ds G_{23} = \left(\epsilon^2 - 2 \epsilon^3 t^2\right)\p_t - 2 x \epsilon^3 t \p_x ,&
\ds G_{24} = \left(2 \epsilon^2 t - \frac{4}{3} \epsilon^3 t^3 \right)\p_t + \left( x
\epsilon^2 - 2 x t^2 \epsilon^3\right)\p_x,
\vspace{2mm}\\
\multicolumn{2}{l}{\ds G_{25} = \left( 2 \epsilon t - \frac{4}{3} \epsilon^2 t^3 +
 \frac{4}{15} \epsilon^3 t^5\right)\p_t
 -\left(2 x \epsilon^2 t^2 -\frac{2}{3} x \epsilon^3 t^4\right) \p_x,}
\vspace{2mm}\\
\multicolumn{2}{l}{\ds G_{26} = \left( t - \frac{2}{3} \epsilon t^3 + \frac{2}{15}
\epsilon^2 t^5 - \frac{4}{315} \epsilon^3 t^7\right)\p_t
+ \left(\frac{1}{3} x \epsilon^2 t^4 - x \epsilon t^2 - \frac{2}{45} x \epsilon^3 t^6\right)\p_x,}
\vspace{2mm}\\
\multicolumn{2}{l}{\ds G_{27}= \left(\epsilon - 2 \epsilon^2 t^2
+ \frac{2}{3} \epsilon^3 t^4\right)\p_t  - \left( 2 x \epsilon^2 t - \frac{4}{3}
x \epsilon^3 t^3\right)\p_x,}
\vspace{2mm}\\
\multicolumn{2}{l}{\ds G_{28} = \left( x \epsilon - \frac 12 x \epsilon^2 t^2 + \frac{1}{4!}
x \epsilon^3 t^4\right)\p_t - \left(x^2 \epsilon^2 t -
\frac{1}{3!} x^2 \epsilon^3 t^3\right)\p_x,}
\vspace{2mm}\\
\multicolumn{2}{l}{\ds G_{29} = \left(\epsilon^2 t^2 - \frac{1}{3} \epsilon^3 t^4
\right)\p_t + \left( x \epsilon^2 t - \frac{2}{3} x \epsilon^3 t^3\right)\p_x,}
\vspace{2mm}\\
\multicolumn{2}{l}{\ds G_{30} = \left(\epsilon t
- \frac{1}{3!} \epsilon^2 t^3 + \frac{1}{5!} \epsilon^3 t^5\right)\p_x,}
\vspace{2mm}\\
\multicolumn{2}{l}{\ds G_{31} = \left(x \epsilon t - \frac{1}{3!} x \epsilon^2 t^3
+ \frac{1}{5!} x \epsilon^3 t^5\right)\p_t
+ \left(x^2 \epsilon - \frac{1}{2!} x^2 \epsilon^2 t^2 + \frac{1}{4!} x^2 \epsilon^3 t^4
\right)\p_x,}
\vspace{2mm}\\
G_{32} = \epsilon^3 x \p_t. &
\ea\nonumber\\ \label{144}
\end{eqnarray}

We observe that there are sixteen approximate symmetries at $O(\epsilon)$,
twenty-four at $O\left(\epsilon^2\right)$ and thirty-two at $O\left(\epsilon^3\right)$.
 On closer
inspection we see that there are different categories of approximate symmetry
which would not become apparent if we worked only to $O(\epsilon)$.  The
homogeneity symmetry, $x\p/\p x$, which is common to both perturbed and
unperturbed equations occurs at all orders of $\epsilon$.  Other symmetries
occur at levels of increasing approximation, for example $G_3$, in which
one sees the r\^ole of the perturbation parameter bringing additional terms
of a MacLaurin expansion.  We recognise the initial terms of the expected
sine and cosine series.  Then also we see, again with the example of $G_3$ at
$O(\epsilon)$, that $G_{12}$ is simply $\epsilon$ times the first order $G_3$.  Hence
$G_3$ is a true
approximate symmetry in contrast with, say, $G_{12}$ which is simply a truncation
symmetry.  At $O(\epsilon)$ one cannot distinguish between the natures of the two
types of symmetry.  Consequently we can recognise three types of ``approximate''
symmetry.  The first category comprises those symmetries which are common to
both perturbed and unperturbed equations and which occur at all orders of the
perturbation parameter.  The second class are those symmetries which are
truely approximate as one can see the increasing level of approximation
with increasing powers of $\epsilon$.
They also occur at all levels of approximation.  The final category
are those which exist by virtue of the power at which the approximation was truncated.
 They can also have expansions of a form similar to the second category,
but are distinguished by never having a term which is independent of~$\epsilon$.
The number of approximate symmetries increases by eight with each additional
order of approximation.

\subsection{The Ermakov--Pinney equation}
The Ermakov--Pinney equation~\cite{ermakov,pinney} is one of those equations
which appears in divers areas and is intimately associated with the~$sl(2,R)$
subalgebra which is the characteristic of all scalar
$n$th order ordinary differential equations of maximal
symmetry~\cite{leach1}.  We write the equation as
\be
\ddot{x}+\omega^2x = \frac{\epsilon}{x^3} \label{3}
\ee
which has the implication of a perturbation from the equation for a simple
harmonic oscillator.  We note that both $\omega$
and $\epsilon$ can be scaled to any
numbers by rescaling of the variables~$x$ and~$t$.  The Ermakov--Pinney equation
has three exact symmetries so that here we are looking at the effect of a
decrease in exact symmetries from eight to three.  Again we
list the approximate symmetries at $O(\epsilon)$, $O\left(\epsilon^2\right)$ and
$O\left(\epsilon^3\right)$.  They are listed in (\ref{4}).

The pattern in (\ref{4}) is perhaps easier to discern than in
(\ref{122}), (\ref{133}) and (\ref{144}).
The first three approximate symmetries, $G_1$, $G_2$
and $G_3$, are repeated at $O(\epsilon)$,
$O\left(\epsilon^2\right)$ and $O\left(\epsilon^3\right)$ as the
order of the approximation is increased and account for the first
twelve approximate symmetries (where appropriate).  These
symmetries, with the algebra $sl(2,R)$, are the symmetries common
to both the equation for the simple harmonic oscillator and that
for the Ermakov--Pinney equation.  The remaining symmetries are
all at the maximum power of $\epsilon$ applicable and so are a
consequence of the truncation of the perturbation expansion.  They
represent the remnants of the homogeneity symmetry ($G_{15}$ and
the solution symmetries ($G_{16}$ and $G_{17}$) of the original
simple harmonic oscillator.  The number of approximate symmetries
increases by three at each additional order of approximation.

{\small
\begin{eqnarray}\hspace{-9mm}
\begin{array}{lll}
O(\epsilon) & O\left(\epsilon^2\right) & O\left(\epsilon^3\right)\\
G_1 = \p_t&G_1 = \p_t&G _1 = \p_t\\
G_2 = \sin2\omega t\p_t&G_2 = \sin2\omega t\p_t&
G_2 = \sin2\omega t\p_t+\omega x\cos2\omega t\p_x\\
+ \omega x\cos2\omega t\p_x &+\omega x\cos2\omega t\p_x&
 \\
G_3 = \cos2\omega t\p_t&G_3 = \cos2\omega t\p_t-\omega x\sin2\omega t\p_x&
G_3 = \cos2\omega t\p_t-\omega x\sin2\omega t\p_x\\
-\omega x\sin2\omega t\p_x&&\\
G_4 = \epsilon \p_t&G_4 = \epsilon \p_t&G_4 = \epsilon \p_t\\
G_5 = \epsilon\sin2\omega t\p_t&G_5 = \epsilon\left[\sin2\omega t\p_t+\omega x\cos2\omega t\p_x\right]&
G_5 = \epsilon\left[\sin2\omega t\p_t+\omega x\cos2\omega t\p_x\right]\\
+\epsilon\omega x\cos2\omega t\p_x&&\\
G_6 = \epsilon\cos2\omega t\p_t&
G_6 = \epsilon\left[ \cos2\omega t\p_t-\omega x\sin2\omega t\p_x\right]&
G_6 = \epsilon\left[ \cos2\omega t\p_t-\omega x\sin2\omega t\p_x\right]\\
-\epsilon\omega x\sin2\omega t\p_x&&\\
&G_7 = \epsilon^2 \p_t&G_7 = \epsilon^2 \p_t\\
&G_8 = \epsilon^2\sin2\omega t\p_t + \epsilon^2 \omega x\cos2\omega t\p_x&
G_8 = \epsilon^2\left[\sin2\omega t\p_t+\omega x\cos2\omega t\p_x\right]\\
&G_9 = \epsilon^2\left[ \cos2\omega t\p_t-\omega x\sin2\omega t\p_x\right]&
G_9 = \epsilon^2\left[ \cos2\omega t\p_t-\omega x\sin2\omega t\p_x\right]\\
&&G_{10} = \epsilon^3\p_t\\
&&G_{11} = \epsilon^3\left[ \sin2\omega t\p_t+\omega x\cos2\omega t\p_x\right]\\
&&G_{12} = \epsilon^3\left[\cos2\omega t\p_t-\omega x\sin2\omega t\p_x\right]\\
G_{13} = \epsilon x\p_x&G_{13} = \epsilon^2x\p_x&G_{13} = \epsilon^3 x\p_x\\
G_{14} = \epsilon \sin\omega t\p_x&G_{14} = \epsilon^2 \sin\omega t\p_x&
G_{14} = \epsilon^3 \sin\omega t\p_x\\
G_{15} = \epsilon \cos\omega t\p_x&G_{15} = \epsilon^2 \cos\omega t\p_x&G_{15} = \epsilon^3
\cos\omega t\p_x\\
G_{16} = \epsilon x\sin\omega t\p_t&
G_{16} = \epsilon^2\left[ x\sin\omega t\p_t+\omega x^2\cos\omega t\p_x\right]&
G_{16} = \epsilon^3\left[ x\sin\omega t\p_t+\omega x^2\cos\omega t\p_x\right]\\
+\epsilon\omega x^2\cos\omega t\p_x&&\\
G_{17} = \epsilon x\cos\omega t\p_t&
G_{17} = \epsilon^2\left[ x\cos\omega t\p_t-\omega x^2\sin\omega t\p_x\right]&
G_{17} = \epsilon^3\left[ x\cos\omega t\p_t-\omega x^2\sin\omega t\p_x\right].\\
- \epsilon \omega x^2\sin\omega t\p_x& &\\
\end{array} \hspace{-10mm}\nonumber \\ \label{4}
\end{eqnarray}}

\subsection{Emden--Fowler equation}
We consider the Emden--Fowler equation
\be
\ddot{x} = \epsilon x^2, \label{5}
\ee
i.e. as a perturbation of the free particle equation.  Again the parameter can
be set at any value by rescaling.  We find the approximate symmetries as are
listed in (\ref{6}).

The symmetries $G_1$ and $G_2$ are repeated at higher orders of approximation
and so are common to both equations.  In the normal listing of the Lie point
symmetries of the free particle one would have the two symmetries
\be
X_1 = x\p_x \qquad \mbox{and}\qquad X_2 = t\p_t+\frac 12 x\p_x, \label{7}
\ee
the homogeneity symmetry and the middle element of the $sl(2,R)$ triplet in the
form appropriate to a second order ordinary
differential equation~\cite{leach1}, but any linear combination of $X_1$
and $X_2$ is also a symmetry.  The combination which gives $G_2$ is common to
both equations and so is maintained at all orders of $\epsilon$ whereas the
combinations which give $G_9$ and $G_{12}$ are not inherited, but are simply a
consequence of the order of the approximate symmetry.  We also observe that the
perturbations of $G_1$ occur to the order of $\epsilon$ being used in the
approximate symmetry whereas those of $G_2$ are at one order less.  This is
because the differential equation, (\ref{5}), is used in the calculation of the
second approximate symmetry but not the first which simply annihilates the
equation.

{\small
\begin{eqnarray}\hspace*{-9mm}
\begin{array}{lll}
O(\epsilon) & O\left(\epsilon^2\right) & O\left(\epsilon^3\right)\\
G_1 = \p_t&G_1 = \p_t&G _1 = \p_t\\
G_2 = t\p_t-2x\p_x&G_2 = t\p_t-2x\p_x&G_2 = t\p_t-2x\p_x\\
G_3 = \epsilon \p_t&G_3 = \epsilon \p_t&G_3 = \epsilon \p_t\\
&G_4 = \epsilon\left[ t\p_t-2x\p_x\right]&G_4 = \epsilon\left[ t\p_t-2x\p_x\right]\\
&G_5 = \epsilon^2 \p_t&G_5 = \epsilon^2 \p_t\\
&&G_6 = \epsilon^2\left[ t\p_t-2x\p_x\right]\\
&&G_7 = \epsilon^3 \p_t\\
G_8 = \epsilon\p_x&G_8 = \epsilon^2\p_x&G_8 = \epsilon^3\p_x\\
G_9 = \epsilon t\p_t&
G_9 = \epsilon^2 t\p_t&
G_9 = \epsilon^3 t\p_t\\
G_{10} = \epsilon x\p_t&G_{10} = \epsilon^2 x\p_t&G_{10} = \epsilon^3 x\p_t\\
G_{11} = \epsilon t\p_x&G_{11} = \epsilon^2 t\p_x&G_{11} = \epsilon^3 t\p_x\\
G_{12} = \epsilon x\p_x&G_{12} = \epsilon^2 x\p_x&G_{12} = \epsilon^3 x\p_x\\
G_{13} = \epsilon\left[ tx\p_t+ x^2\p_x\right]&
G_{13} = \epsilon^2\left[ tx\p_t+ x^2\p_x\right]&
G_{13} = \epsilon^3\left[ tx\p_t+ x^2\p_x\right]\\
G_{14} = \epsilon\left[ t^2\p_t +  tx\p_x\right]&
G_{14} = \epsilon^2\left[ t^2\p_t + tx\p_x\right]&
G_{14} = \epsilon^3\left[ t^2\p_t + tx\p_x\right]\\
G_{15} = \epsilon t^4\p_t&
G_{15} = \epsilon\left[\epsilon t^4\p_t+ \left(6t+2\epsilon t^3x\right)\p_x\right]
&G_{15} = \epsilon^2\left[ t^4\p_t+ \left(6t+2\epsilon t^3x\right)\p_x\right]\\
+ \left(6t+2\epsilon t^3x\right)\p_x&&\\
G_{16} = 2\epsilon t^3\p_t&
G_{16} = \epsilon\left[ 2\epsilon t^3\p_t+ 3\left(1+\epsilon t^2x\right)\p_x\right]
&G_{16} = \epsilon^2\left[ 2\epsilon t^3\p_t+ 3\left(1+\epsilon t^2x\right)\p_x\right].\\
+ 3\left(1+\epsilon t^2x\right)\p_x&&\\
\end{array} \hspace{-10mm}\nonumber\\ \label{6}
\end{eqnarray}}

\setcounter{equation}{0}

\section{The Noether symmetries and approximate\\
 first integrals/invariants}
\subsection{The simple harmonic oscillator}
The simple harmonic oscillator has the Lagrangian
\be
L = \frac 12 \left[\dot{x}^2-\epsilon x^2\right]. \label{8}
\ee
Given the observations made in the previous section we know the pattern of the
behaviour of the symmetries and so we will simply list the results for the
first two orders of~$\epsilon$.  We consider point symmetries only.  Because of the
direct relationship between symmetry
and first integral we list the $G$s, $f$s and $I$s together.  The results are
given in (\ref{9}) and (\ref{10}).

The only Noether symmetry of the free particle which is inherited by the simple
harmonic oscillator is $G_1$.  Since both Lagrangians have five Noether point
symmetries, at $O(\epsilon)$ we see that there are two symmetries due to the
inheritance of the exact symmetry $G_1$, four approximate symmetries of the
free particle which persist as truncation symmetries and four genuine
approximation symmetries of the simple harmonic oscillator.  We see the
beginnings of the MacLaurin expansions for the sine and cosine functions.  The
behaviour of the integrals reflects these distinctions.  At $O\left(\epsilon^2\right)$
we gain
an extra five symmetries.  One of these is due to the inherited symmetry and
the other four because the four approximate symmetries of the simple harmonic
oscillator at order one will have the same behaviour as the inherited symmetry
at order two.  The distinction between the two classes is that for the
inherited symmetry all occurrences are simply multiples of the zeroth order
symmetry by powers of $\epsilon$ whereas the basic approximation symmetries
increase the degree of approximation with increasing powers of $\epsilon$ and the
others are multiple of approximate symmetries at lower order.  The truncation
symmetries do not multiply, but simply persist at the order of the
approximation being used.

To
the first
order in $\epsilon$ we have
{\small
\begin{eqnarray}\hspace*{-9mm}
\begin{array}{lll}
\hline
\mbox{\rm Symmetry}&\mbox{Gauge function}&\mbox{Invariant}\\
\hline
G_1 = \p_t&f_1 = 0&I_1 = \frac{1}{2}\left(\dot{x}^2+\epsilon x^2\right)\\
G_2 = \epsilon\p_t&f_2 = 0& I_2 = \frac{1}{2}\epsilon\dot{x}^2\\
G_3 = \left[ 2t- \frac{1}{3!}\epsilon(2t)^3\right]\p_t&
f_3 = -2\epsilon tx^2&I_3 =
\dot{x}\left(t\dot{x}-x\right)-\epsilon t\left(x^2-2tx\dot{x}+\frac{2}{3}t^2\dot{x}^2\right)\\
+ x\left[ 2t-\frac{1}{2!}\epsilon(2t)^2\right]\p_x&&\\
G_4 = \epsilon\left(2t\p_t+x\p_x\right)&f_4 = 0&I_4 = \epsilon\dot{x}\left(t\dot{x}-x\right)\\
G_5 = \left(1-\frac{1}{2!}\epsilon t^2\right)\p_x&f_5 = -\epsilon tx&I_5 = \left(1-\frac{1}{2!}\epsilon t^2\right)\dot{x} + \epsilon tx\\
G_6 = \epsilon\p_x&f_6 = 0&I_6 = \epsilon \dot{x}\\
G_7 = \left(t-\frac{1}{3!}\epsilon t^3\right)\p_x&f_7 = \left(1-\frac{1}{2!}\epsilon t^2\right)x&I_7 = \left(1-\frac{1}{2!}\epsilon t^2\right)x
- \left(t-\frac{1}{3!}\epsilon t^3\right)\dot{x}\\
G_8 = \epsilon t\p_x&f_8 = \epsilon x&I_8 = \epsilon\left(x-t\dot{x}\right)\\
G_9 = \left[ \frac{1}{2!}(2t)^2-\frac{1}{4!}\epsilon(2t)^4\right]\p_t&
f_9 =
\left(1-\frac{1}{2!}\epsilon(2t)^2\right)x^2&I_9 = \left(x-t\dot{x}\right)^2-\frac{1}{3}\epsilon t^2\left(3x-t\dot{x}\right)\left(x-t\dot{x}\right)\\
+x\left[ 2t-\frac{1}{3!}\epsilon(2t)^3\right]\p_x&&\\
G_{10} = \frac{1}{2!}\epsilon\left((2t)^2\p_t+x(2t)\p_x\right)&f_{10} = \epsilon x^2&I_{10} =
\epsilon\left(x-t\dot{x}\right)^2\\
\hline
\end{array}  \hspace{-10mm}\nonumber\\ \label{9}
\end{eqnarray}}
\noindent and to the second order in $\epsilon$ we have
{\small
\begin{eqnarray}\hspace*{-9mm}
\begin{array}{lll}
\hline
\mbox{\rm Symmetry}&\mbox{Gauge function}&\mbox{Invariant}\\
\hline
G_1 = \p_t&f_1 = 0&I_1 = \frac{1}{2}\left(\dot{x}^2+\epsilon x^2\right)\\
G_2 = \epsilon\p_t&f_2 = 0& I_2 = \frac{1}{2}\epsilon\left(\dot{x}^2+\epsilon x^2\right)\\
G_3 = \epsilon^2\p_t&f_3 = 0&I_3 = \frac{1}{2}\epsilon\dot{x}^2\\
G_4 = \left[ 2t- \frac{1}{3!}\epsilon(2t)^3 + \frac{1}{5!}\epsilon^2(2t)^5\right]\p_t&
f_4 = -\epsilon x^2\left[ 2t-\frac{1}{3!}\epsilon(2t)^3\right]&
I_4 = \dot{x}\left(t\dot{x}-x\right)\\
+ x\left[ 2t-\frac{1}{2!}\epsilon(2t)^2+\frac{1}{4!}\epsilon^2(2t)^4\right]\p_x&&
 - \frac{1}{3}\epsilon t\left(3\dot{x}^2-6tx\dot{x}+2t^2\dot{x}^2\right)\\
&&+\frac{2}{15}\epsilon^2t^2\left(5x^2- 5tx\dot{x}+3t^2\dot{x}^2\right)\\
G_5 = \epsilon\left[ 2t- \frac{1}{3!}\epsilon(2t)^3\right]\p_t&
f_5 = -2\epsilon^2 tx^2&I_5 =
\epsilon\dot{x}\left(t\dot{x}-x\right)\\
 + \epsilon x\left[ 2t-\frac{1}{2!}\epsilon(2t)^2 \right]\p_x& &
-\epsilon^2 t\left(x^2-2tx\dot{x}+\frac{2}{3}t^2\dot{x}^2\right)\\
G_6 = \epsilon^2\left(2t\p_t+x\p_x\right)&f_6 = 0&I_6 = \epsilon^2\dot{x}\left(t\dot{x}-x\right)\\
G_7 = \left(1-\frac{1}{2!}\epsilon t^2+\frac{1}{4!}\epsilon^2t^4\right)\p_x&f_7 = -\epsilon x\left(t-\frac{1}{3!}\epsilon
t^3\right)
&I_7 = \left(1-\frac{1}{2!}\epsilon t^2+\frac{1}{4!}\epsilon^2t^4\right)\dot{x}\\
&& + \epsilon x\left(t-\frac{1}{3!}\epsilon
t^3\right)\\
G_8 = \epsilon\left(1-\frac{1}{2!}\epsilon t^2\right)\p_x&f_5 = -\epsilon^2 tx&I_5 = \epsilon\left[\left(1-\frac{1}{2!}
\epsilon t^2\right)\dot{x}
+ \epsilon tx\right]\\
G_9 = \epsilon^2\p_x&f_9 = 0&I_9 = \epsilon^2 \dot{x}\\
G_{10} = \left(t-\frac{1}{3!}\epsilon t^3+\frac{1}{5!}\epsilon^2t^5\right)\p_x&f_{10} = \left(1-\frac{1}{2!}\epsilon
t^2+\frac{1}{4!}\epsilon^2t^4\right)x&I_{10} = \left(1-\frac{1}{2!}\epsilon t^2+\frac{1}{4!}\epsilon^2t^4\right)x\\
&&- \left(t-\frac{1}{3!}\epsilon t^3+\frac{1}{5!}\epsilon^2t^5\right)\dot{x}\\
G_{11} = \epsilon\left(t-\frac{1}{3!}\epsilon t^3\right)\p_x&f_{11} = \epsilon\left(1-\frac{1}{2!}\epsilon t^2\right)x
&I_{11} = \epsilon\left(1-\frac{1}{2!}\epsilon t^2\right)x\\
&& - \epsilon\left(t-\frac{1}{3!}\epsilon t^3\right)\dot{x}\\
G_{12} = \epsilon^2 t\p_x&f_{12} = \epsilon^2 x&I_{12} = \epsilon^2\left(x-t\dot{x}\right)\\
G_{13} = \left[ \frac{1}{2!}(2t)^2-\frac{1}{4!}\epsilon(2t)^4+\frac{1}{6!}\epsilon^2t^6\right]\p_t&
f_{13} =
\left(1-\frac{1}{2!}\epsilon(2t)^2\right)^2
&I_{13} = \left(x-t\dot{x}\right)^2-\\
+x\left[ 2t-\frac{1}{3!}\epsilon(2t)^3+\frac{1}{5!}\epsilon^2t^5\right]\p_x&
+\frac{1}{4!}\epsilon^2(2t)^4 x^2&
\frac{1}{3}\epsilon t^2
\left(3x-t\dot{x}\right)\left(x-t\dot{x}\right)\\
&&+\frac{1}{15}\epsilon^2t^4\left(5x^2-4tx\dot{x}+2t^2
\dot{x}^2\right)\\
G_{14} = \epsilon\left( \frac{1}{2!}(2t)^2-\frac{1}{4!}\epsilon(2t)^4\right)\p_t&
f_{14} = \epsilon\left(1-\frac{1}{2!}\epsilon(2t)^2\right)x^2&I_{14} = \epsilon\left(x-t\dot{x}\right)^2\\
+\epsilon x\left(2t-\frac{1}{3!} \epsilon(2t)^3\right)\p_x&&
-\frac{1}{3}\epsilon^2 t^2\left(3x-t\dot{x}\right)\left(x-t\dot{x}\right)\\
G_{15} = \frac{1}{2!}\epsilon^2\left((2t)^2\p_t+x(2t)\p_x\right)&f_{15} = \epsilon^2 x^2&I_{15} =
\epsilon^2\left(x-t\dot{x}\right)^2\\
\hline
\end{array}  \hspace{-10mm}\nonumber\\ \label{10}
\end{eqnarray}}

\subsection{Ermakov--Pinney system}
The Lagrangian for the Ermakov--Pinney system is
\be
L = \frac{1}{2}\left[\dot{x}^2-\omega^2x^2-\frac{\epsilon}{x^2}\right]. \label{11}
\ee
To the first order in $\epsilon$ we have
{\small
\begin{eqnarray}\hspace*{-9mm}
\begin{array}{lll}
\hline
\mbox{\rm Symmetry}&\mbox{Gauge function}&\mbox{Invariant}\\
\hline
G_1 = \p_t&f_1 = 0&I_1 = \frac{1}{2}\left(\dot{x}^2+\omega^2 x^2+
\ds{\frac{\epsilon}{x^2}}\right)\\
G_2 = \epsilon\p_t&f_2 = 0&I_2 = \frac{1}{2}\epsilon\left(\dot{x}^2+\omega^2 x^2\right)\\
G_3 = \sin 2\omega t\p_t + \omega x\cos 2\omega t\p_x&f_3 = -\omega^2x^2\sin 2\omega t&
I_3 = \frac{1}{2}\left(\dot{x}^2-\omega^2x^2+\ds{\frac{\epsilon}{x^2}}\right)\sin 2\omega t\\
&& -\omega x\dot{x}\cos 2\omega t\\
G_4 = \cos 2\omega t\p_t - \omega x\sin 2\omega t\p_x&f_4 = \omega^2x^2\cos 2\omega t&
I_4 = \frac{1}{2}\left(\dot{x}^2-\omega^2x^2+\ds{\frac{\epsilon}{x^2}}\right)\cos 2\omega t\\
&& +\omega x\dot{x}\sin 2\omega
t\\
G_5 = \epsilon\left[\sin 2\omega t\p_t + \omega x\cos 2\omega t\p_x\right]&f_5 = -\epsilon\omega^2x^2\sin
2\omega t&
I_5 = \epsilon\frac{1}{2}\left(\dot{x}^2-\omega^2x^2\right)\sin 2\omega t\\
&& -\epsilon \omega x\dot{x}\cos 2\omega t\\
G_6 = \epsilon\left[\cos 2\omega t\p_t - \omega x\sin 2\omega t\p_x\right]&f_6 = \epsilon\omega^2x^2\cos
2\omega t&I_6 = \epsilon \frac{1}{2}\left(\dot{x}^2-\omega^2x^2\right)\cos 2\omega t\\
& & + \epsilon \omega x\dot{x}\sin 2\omega t\\
G_7 = \epsilon\sin\omega t\p_x&f_7 = \epsilon\omega x\cos\omega t&I_7 = \epsilon\left(\dot{x}\sin\omega t -
\omega x\cos\omega t\right)\\
G_8 = \epsilon\cos\omega t\p_x&f_8 = -\epsilon\omega x\sin\omega t&I_8 = \epsilon\left(\dot{x}\cos\omega t +
\omega x\sin\omega t\right)\\
\hline
\end{array}  \hspace{-10mm}\nonumber\\ \label{12}
\end{eqnarray}}
\noindent and to the second order in $\epsilon$
{\small
\begin{eqnarray}\hspace*{-9mm}
\begin{array}{lll}
\hline
\mbox{\rm Symmetry}&\mbox{Gauge function}&\mbox{Invariant}\\
\hline
G_1 = \p_t&f_1 = 0&I_1 = \frac{1}{2}\left(\dot{x}^2+\omega^2 x^2+\ds{\frac{\epsilon}{x^2}}\right)\\
G_2 = \epsilon\p_t&f_2 = 0&I_2 = \frac{1}{2}\epsilon\left(\dot{x}^2+\omega^2 x^2+\ds{\frac{\epsilon}{x^2}}\right)\\
G_3 = \epsilon^2\p_t&f_3 = 0&I_3 = \frac{1}{2}\epsilon^2\left(\dot{x}^2+\omega^2 x^2\right)\\
G_4 = \sin 2\omega t\p_t + \omega x\cos 2\omega t\p_x&f_4 = -\omega^2x^2\sin 2\omega t&
I_4 = \frac{1}{2}\left(\dot{x}^2-\omega^2x^2+\ds{\frac{\epsilon}{x^2}}\right)\sin 2\omega t\\
&& -\omega x\dot{x}\cos 2\omega t\\
G_5 = \cos 2\omega t\p_t - \omega x\sin 2\omega t\p_x&f_5 = \omega^2x^2\cos 2\omega t&
I_5 = \frac{1}{2}\left(\dot{x}^2-\omega^2x^2+\ds{\frac{\epsilon}{x^2}}\right)\cos 2\omega t\\
&& +\omega x\dot{x}\sin 2\omega t\\
G_6 = \epsilon\left[\sin 2\omega t\p_t + \omega x\cos 2\omega t\p_x\right]&f_6 = -\epsilon\omega^2x^2\sin
2\omega t&
I_6 = \frac{1}{2} \epsilon \left(\dot{x}^2-\omega^2x^2+\ds{\frac{\epsilon}{x^2}}\right)\sin 2\omega t\\
&& -\epsilon \omega x\dot{x}\cos 2\omega t\\
G_7 = \epsilon\left[\cos 2\omega t\p_t - \omega x\sin 2\omega t\p_x\right]&f_7 = \epsilon\omega^2x^2\cos 2\omega
t&I_7 = \frac{1}{2} \epsilon \left(\dot{x}^2-\omega^2x^2+\ds{\frac{\epsilon}{x^2}}\right)\cos 2\omega t\\
&& + \epsilon \omega x\dot{x}\sin 2\omega t\\
G_8 = \epsilon^2\sin\omega t\p_x&f_8 = \epsilon^2\omega x\cos\omega t&I_8 = \epsilon^2\left(\dot{x}\sin\omega
t - \omega x\cos\omega t\right)\\
G_9 = \epsilon^2\cos\omega t\p_x&f_8 = -\epsilon^2\omega x\sin\omega t&I_8 = \epsilon^2\left(\dot{x}\cos\omega t +
\omega x\sin\omega t\right)\\
G_{10} = \epsilon^2\left[\sin 2\omega t\p_t + \omega x\cos 2\omega t\p_x\right]&f_{10} =
-\epsilon^2\omega^2x^2\sin 2\omega t&
I_{10} = \frac{1}{2} \epsilon^2\left(\dot{x}^2-\omega^2x^2\right)\sin 2\omega t\\
& &  -\epsilon^2 \omega x\dot{x}\cos 2\omega t\\
G_{11} = \epsilon^2\left[\cos 2\omega t\p_t - \omega x\sin 2\omega t\p_x\right]&f_{11} =
\epsilon\omega^2x^2\cos 2\omega
t&I_{11} = \frac{1}{2} \epsilon^2\left(\dot{x}^2-\omega^2x^2\right)\cos 2\omega t\\
& & + \epsilon^2\omega x\dot{x}\sin 2\omega t.\\
\hline
\end{array} \hspace{-10mm}\nonumber\\ \label{13}
\end{eqnarray}}

We observe that the three symmetries, $G_1$, $G_2$ and $G_3$, of (\ref{4}) are
common Lie and Noether symmetries of the equation for the oscillator and the
Ermakov--Pinney equation.  They constitute the algebra $sl(2,R)$ which is the
symmetry algebra for the latter and a subalgebra of the $sl(3,R)$ algebra of
the former.  The solution symmetries are preserved by the order of truncation.
 The other three approximate Lie symmetries are not approximate
Noether symmetries.  The only serious first integrals/invariants correspond
to these three symmetries.  They are exact and not approximate.

\subsection{Emden--Fowler Lagrangian}
The Lagrangian for the Emden--Fowler equation (\ref{5}) is
\be
L = \frac{1}{2}\dot{x}^2+\frac 13 \epsilon x^3. \label{14}
\ee
To the first order in $\epsilon$ we have
{\small
\begin{eqnarray}\hspace*{-9mm}
\begin{array}{lll}
\hline
\mbox{\rm Symmetry}&\mbox{Gauge function}&\mbox{Invariant}\\
\hline
G_1 = \p_t&f_1 = 0 &I_1 = \frac{1}{2}\dot{x}^2-\frac 13 \epsilon x^3\\
G_2 = \epsilon\p_t&f_2 = 0&I_2 = \frac{1}{2}\epsilon\dot{x}^2\\
G_3 = \epsilon\p_x&f_3 = 0&I_3 = \epsilon\dot{x}\\
G_4 = \epsilon t\p_x&f_4 = 0&I_4 = \epsilon(x-\dot{x}t)\\
G_5 = \epsilon\left[ 2t\p_t+x\p_x\right]&f_5 = 0&I_5 = \epsilon\dot{x}(x-\dot{x}t) \\
G_6 = \epsilon\left[ t^2\p_t+tx\p_x\right]&f_6 = \frac{1}{2}\epsilon x^2&I_6 = \frac{1}{2}\epsilon(x-t\dot{x})^2\\
G_7 = \frac{1}{3!}\epsilon t^4\p_t+\left(t+\frac{1}{3}\epsilon t^3x\right)\p_x&f_7 =\frac{1}{2}\epsilon t^2x^2 &I_7 =
t\dot{x}-\epsilon\left(\frac{1}{12}t^4-\frac{1}{3} t^3x\dot{x}+\frac{1}{2} t^2x^2\right)\\
G_8 = \frac{2}{3}\epsilon t^3\p_t+\left(1+\epsilon t^2x\right)\p_x&f_8 = \epsilon tx^2&I_8 = \dot{x} -
\epsilon\left(tx^2-t^2x\dot{x}+\frac{1}{3} t^3\dot{x}^2\right)\\
\hline
\end{array}  \hspace{-10mm}\nonumber\\ \label{15}
\end{eqnarray}}
\noindent and to the second order in $\epsilon$ we have
{\small
\begin{eqnarray}\hspace*{-9mm}
\begin{array}{lll}
\hline
\mbox{\rm Symmetry}&\mbox{Gauge function}&\mbox{Invariant}\\
\hline
G_1 = \p_t&f_1 = 0 &I_1 = \frac{1}{2}\dot{x}^2-\frac{1}{3}\epsilon x^3\\
G_2 = \epsilon\p_t&f_2 = 0&I_2 = \epsilon\left(\frac{1}{2}\dot{x}^2-\frac{1}{3}\epsilon x^3\right)\\
G_3 = \epsilon^2\p_t&f_3 = 0&I_3 = \frac{1}{2}\epsilon^2\dot{x}^2\\
G_4 = \epsilon^2\p_x&f_3 = 0&I_4 = \epsilon^2\dot{x}\\
G_5 = \epsilon^2 t\p_x&f_5 = 0&I_5 = \epsilon^2(x-\dot{x}t)\\
G_6 = \epsilon^2\left[ 2t\p_t+x\p_x\right]&f_6 = 0&I_6 = \epsilon^2\dot{x}(x-\dot{x}t) \\
G_7 = \epsilon^2\left[ t^2\p_t+tx\p_x\right]&f_7 = \frac{1}{2}\epsilon^2 x^2&I_7 = \frac{1}{2}\epsilon^2(x-t\dot{x})^2\\
G_8 = \epsilon\left[\frac{1}{3!}\epsilon t^4\p_t+\left(t+\frac{1}{3}\epsilon t^3x\right)\p_x\right]&f_8 = \frac{1}{2}\epsilon^2 t^2x^2&
I_8 = \epsilon\left[ t\dot{x}-\epsilon\left(\frac{1}{12}t^4-\frac{1}{3} t^3x\dot{x}+\frac{1}{2} t^2x^2\right)\right]\\
G_9 = \epsilon\left[\frac{2}{3}\epsilon t^3\p_t+\left(1+\epsilon t^2x\right)\p_x\right]&f_9 = \epsilon tx^2&
I_9 = \epsilon\left[ \dot{x} - \epsilon\left(tx^2-t^2x\dot{x}+\frac{1}{3} t^3\dot{x}^2\right)\right]\\
\hline
\end{array}  \hspace{-10mm}\nonumber\\ \label{16}
\end{eqnarray}}
The only Noether symmetry which produces a useful integral is $G_1$.  Had one
treated only to $O(\epsilon)$, which is the implication of the one of the few
texts giving a treatment of approximate symmetries \cite{ibragimov}, one may
have become enthusiastic about the integrals produced by $G_7$ and $G_8$.
Indeed they appear to be classic examples of a perturbation solution.  However,
we see from the $O\left(\epsilon^2\right)$
results that these symmetries are simply the
result of truncation since the order of $\epsilon$ increases with the order of the
approximation.  We remark that the results for this Lagrangian are not typical
of Lagrangians of the general Emden--Fowler equation.  It is well-known that the
cases $n=2$ and $n=-3$ differ from the general case~\cite{leach3}.
 Were we to use another value of~$n$,
the number of approximate symmetries would differ and the
symmetries, $G_7$ and $G_8$, at the first order would not occur.
We note that the Lie symmetry, $G_2$, of this Emden--Fowler
equation is not a Noether symmetry.

\setcounter{equation}{0}

\section{Lie integrals}
The attraction of Noether's Theorem is that the procedure for
obtaining a first integral/invariant is trivial once the symmetry
and gauge function have been determined, a process which is
simultaneous.  Of course there is a price to pay and that is the
close relationship between the dependence of the coefficient
functions on the derivatives and that of the
integral/invariant~\cite{sc81}. One must look to generalized
symmetries to be assured of finding the requisite number of
invariants to know that a system is integrable.  Apart from the
practical problem of a more complicated calculation there is
always the concern that one may have used the wrong {\it Ansatz}
for the functional dependence on the derivatives in the
coefficient functions.  The Lie method is kinder in that respect,
but the kindness is tempered with the additional burden of not
completely friendly calculations.  For multidimensional systems a
combination of Noether's Theorem coupled with Lie symmetries has
been found to be advantageous, particularly as, in the Lie
approach, one can always insist that the integral be invariant
under more than one symmetry~\cite{Hara}.

We noted at the conclusion of the previous section that the Lie symmetry,
$G_2$, listed in~(\ref{6})
is not a Noether symmetry and so does not have the easy route
to an invariant which Noether's Theorem provides.
We know from the existence of $I_1$ in (\ref{16})
 that the       Emden--Fowler equation (\ref{5})
 is solvable as a consequence of Liouville's theorem~\cite{liouville}.
However, can we look to $G_2$ to provide an invariant which with
$I_1$ will give the solution?  For this we use the Lie method outlined
in \S~1, (\ref{8a}).
 The invariants of $G_2 = t\p/\p t - 2x\p/\p x$
are found from the solution of the associated Lagrange's system
\be \frac{{\rm d} t}{t} = \frac{{\rm d} x}{-2x} = \frac{{\rm d}
\dot{x}}{-3\dot{x}} \label{18} \ee and are \be u =
xt^2\qquad\mbox{and}\qquad v = \dot{x}t^3. \label{19} \ee The
second requirement of (\ref{8a}) is now expressed as \be
\frac{{\rm d}I(u(t,x),v(t,\dot{x}))}{{\rm d}t} = 0 \label{20} \ee
and leads to the second associated Lagrange's system, \be
\frac{{\rm d} u}{v+2u} = \frac{{\rm d} v}{\epsilon u^2+3v}
\label{21} \ee which is an Abel's equation of the second kind and,
like most Abel's equations of either kind, is not responsive to
known methods of integration. Consequently the existence of a
second exact Lie point symmetry does not help in the search for a
second integral.  In terms of the conventional approach one is not
surprised as the Lie Bracket of $G_1$ and~$G_2$ is $[G_1,G_2] =
G_1$ and we are effectively reducing the order by means of the
non-normal subgroup, $G_2$, which is well-known to remove $G_1$ as
a point symmetry of the reduced equation. However, one may use an
unconventional approach to attain our desired objective.  In the
reduction of order we know that the symmetry, $G_1$, becomes
nonlocal.  We know that the type of nonlocality is quite specific.
The point symmetry becomes an exponential nonlocal symmetry and
exponential nonlocal symmetries are known to be sufficient for the
reduction of order~\cite{leach5}.  This is not precisely what we
wish to do here since the second order ordinary differential
equation has already been reduced to a first order ordinary
differential equation and the further reduction to an algebraic
equation, while perhaps of some mathematical interest and
pleasure, does not really further for the solution process. What
we wish to do is to use the nonlocal symmetry to transform the
first order ordinary differential equation to a form such that the
solution of the first order ordinary differential equation
 is reduced to an elementary
procedure.  The reversion of the process of reduction of order to a first
order ordinary differential equation
which can be solved is not normally a trivial task.  Here this is not a concern
as we already have one first integral and that coupled with the invariant we
seek now is all that is required.  (Of course we know that the solution of
this Emden--Fowler equation is expressed in terms of elliptic functions and so
we do not expect to be able to write $x(t)$ in a simple form.)

The symmetry, $G_1$, becomes
\be
X_1 = \exp\left[ -\int\frac{{\rm d} u}{2u+v}\right]\left(2u\frac{\p}{\p u}+3v
\frac{\p}{\p v}\right) \label{22}
\ee
when we reduce the order of (\ref{5})
to the first order ordinary differential equation (\ref{21}).  Were (\ref{22}) a point
symmetry we would seek a
transformation which would make (\ref{21}) autonomous by converting (\ref{22}) to a
simple $\p/\p X$.  However, (\ref{22}) is not a point transformation and we cannot
expect to be able to achieve that much.  We use a modification of the standard
method which seeks a transformation~\cite[p.~110]{bluman}
\be
X = F(u,v),\qquad Y = G(u,v) \label{23a}
\ee
so that the functions $F$ and $G$ satisfy the linear first order
partial differential equations
\be
X_1F=1\qquad\mbox{and}\qquad X_1G=0. \label{23}
\ee
We simply replace the 1 in the first of (\ref{23}) by the exponential part of the
nonlocal symmetry (\ref{22}), so that in the new coordinates $X_1$ will still be
an exponential nonlocal symmetry of the form
\be
\bar{X_1} = \exp\left[ -\int\frac{{\rm d} u}{2u+v}\right]\frac{\p}{\p X}. \label{24a}
\ee
Naturally the integral would be expressed in the new variables when the
relationship is known.  This stratagem means that the exponential term does not
enter into any calculations.  The characteristic for the independent variable is
easily computed to be
\be
w = vu^{-\frac 32}.  \label{24}
\ee
The other characteristic (for the first of (\ref{23})) requires the solution of
\be
0 = {\rm d} F - \frac{{\rm d} u}{2u} \label{25}
\ee
which very easily gives
\be
\zeta  = F - \frac 12\log u. \label{26}
\ee
Under the change of variables
\be
X = vu^{-\frac 32},\qquad Y = \log u^{\frac 12} \label{27}
\ee
(\ref{21}) takes the somewhat more friendly form
\be
0 = \frac{{\rm d}Y}{{\rm d} X} + \frac{X+2\exp[-Y]}{3X^2-2\epsilon} \label{28}
\ee
which is linear in the variable $\exp[Y]$
and we immediately obtain the invariant
\be
J = \mbox{\rm e}^{Y}{3X^2-2\epsilon}^{1/6}+2\int\left(3X^2-2\epsilon\right)^{-5/6}
{\rm d} X. \label{29}
\ee
That the expression for $J$ is complicated is not surprising, but it is a
precise form and maybe gives a different approach to the investigation of
elliptic functions.

The case of the Emden--Fowler equation has not given much joy for the use of
approximate symmetries to obtain approximate integrals/invariants as there
are none which satisfy the criterion of not disappearing with increasing
powers of the small parameter~$\epsilon$.  However, we can illustrate the way
approximate symmetries can give truely approximate integrals/invariants with
the simple harmonic oscillator.  We should not be surprised at this for, if
something does not work for the simple harmonic oscillator, it shall never
work for anything.  One regrets that there are some procedures which work
only for the simple harmonic oscillator and that is the only model upon
which trials are conducted!

\section{Discussion}
We have specified the approximate symmetries for the simple harmonic oscillator
treated as a perturbation of the free particle, the Ermakov--Pinney equation
treated as a perturbation of the equation for the simple harmonic oscillator
and an autonomous Emden--Fowler equation of order two considered as a
perturbation of the free particle. In the case of the simple harmonic
oscillator the number of exact symmetries is eight. The number of approximate
symmetries was found to be sixteen at $O(\epsilon)$, twenty-four at
$O\left(\epsilon^2\right)$ and
thirty-two at $O\left(\epsilon^3\right)$.
The homogeneity symmetry $x \p/\p x$ occurs at all
orders of $\epsilon$. The type of approximate symmetries that
arise can be put into three categories: Those that are common to both perturbed and
unperturbed equations; then the truely approximate symmetries the level of
approximation of which increases with increasing powers of $\epsilon$. The third category
consists of those symmetries in which every term is dependent on $\epsilon$.
The Ermakov--Pinney equation has three exact symmetries instead of eight. When
treated as a perturbation of the equation for the simple harmonic oscillator
it
has eleven approximate symmetries at $O(\epsilon)$, fourteen at
$O\left(\epsilon^2\right)$ and
seventeen at $O\left(\epsilon^3\right)$.
The Ermakov--Pinney equation as a perturbation of the
free particle has twelve approximate symmetries at $O(\epsilon)$, fourteen at
$O\left(\epsilon^2\right)$ and sixteen at $O\left(\epsilon^3\right)$.
Another aspect which has been investigated is that of Noether symmetries and
approximate first integrals (or invariants).

It is important to note that the knowledge of exact
symmetries of a differential equation enables us to reduce it and, if possible,
completely solve it to obtain exact solutions. However, if one seeks for
approximate rather than exact solutions, then approximate symmetries are as
useful as exact ones and can be determined perturbatively. In this paper we
have emphasized more the study of approximate symmetries rather than
approximate solutions.  Finally we should note that in principle there exists
the possibility of the intrusion of secular terms into the expressions for the
approximate symmetries and integrals.  This is not the case with the equations
and integrals treated here.  As applications of approximate symmetries have so
far been at the first order in $\epsilon$, whence the motivation of this paper which
was to look beyond the first order in $\epsilon$, there is no evidence that secular
terms and lack of convergence would present a problem.  The purpose of this
paper was to demonstrate the different types of approximate symmetry available,
a result which is not accessible when approximate symmetries are treated to the
first order only as has been the case in the earlier literature.

\subsection*{Acknowledgements}

PGLL thanks Professor G P Flessas, Dean of the School of Sciences, and
Dr S~Cotsakis, Director of GEODYSYC, for their kind hospitality while this work was
undertaken and the National Research Foundation of South Africa and the
University of Natal for a sabbatical grant.

\label{leach-lastpage}

\begin{thebibliography}{99}
\small
\topsep0mm
\partopsep0mm
\parsep0mm
\itemsep0mm
\bibitem{baikov1}
Baikov V A, Gazizov R K and Ibramigov N H, Approximate Symmetries of Equations
with a Small Parameter, {\it Mat. Sb.}, 1988, V.136, 435--450 ({\it Math. USSR Sb.},
1989, V.64, 427--441).
\bibitem{baikov2}
Baikov V A, Gazizov R K and
Ibramigov N H, Approximate Transformation Groups and
Deformations of Symmetry Lie Algebras, in CRC Handbook of Lie Group
Analysis of Differential Equations, Editor N~H~Ibragimov,
CRC Press, Boca Raton, Florida, 1995, V.3, Ch.2.
\bibitem{bluman}
Bluman G W and Kumei S, Symmetries and Differential Equations,
{\it Appl. Math. Sci.}, Springer-Verlag, New York, 1989, V.81.
\bibitem{Hara}
Cotsakis S, Leach P G L and Pantazi H, Symmetries of Homogeneous
Cosmologies, {\it Grav. {\rm \&} Cosmol.}, 1998, V.4, 314--325.
\bibitem{ermakov}
Ermakov V, Second Order Differential Equations.  Conditions of Complete
Integrability, {\it Univ. Izvestia Kiev, Ser. III}, 1880, V.9, 1--25 (trans. A~O~Harin).
\bibitem{gazizov}
Gazizov R K, Invariants of Approximate Transformation Groups, in Modern
Group Analysis VI, Editors N~H~Ibragimov and F~M~Mahomed,
New Age International Publishers, New Dehli, 1977, 149--158.
\bibitem{leach5}
G\'{e}ronimi C, Feix M R and Leach P G L, Exponential Nonlocal
Symmetries and Nonnormal Reduction of Order,
Preprint: MAPMO, D\'{e}partement
Math\'{e}matique, Universit\'{e} d'Orl\'{e}ans, Orl\'{e}ans la Source, 45067
Orl\'{e}ans Cedex 2, France, 1997.
\bibitem{leach2}
Govinder K S and Leach P G L, The Nature and Uses of Symmetries of Ordinary
Differential Equations, {\it S. Afr. J. Sci.}, 1996, V.92, 23--28.
\bibitem{knn98}
Govinder K S, Heil T G and Uzer T, Approximate Noether Symmetries,
 {\it Phys. Lett. A}, 1998, V.240, 127--131.
\bibitem{head}
Head A K, {\tt LIE}, a PC Program for Lie Analysis of Differential Equations,
{\it Comp. Phys. Comm.}, 1993, V.77, 241--248.
\bibitem{ibragimov}
Ibragimov N H, Elementary Lie Group Analysis and Ordinary Differential
Equations, John Wiley \& Sons Ltd, Chichester, 1999, 231--234.
\bibitem{johnson}
Johnson R, Private Communication, 1979.
\bibitem{Komar1}
Komar A, Asymptotic Covariant Conservation Laws for Gravitational
Rediation, {\it Phys. Rev.}, 1962, V.127, 1411--1418.
\bibitem{Komar2}
Komar A, Positive-Definite Energy Density and Global Consequences
for General Relativity, {\it Phys. Rev.}, 1963, V.129, 1873--1876.
\bibitem{Flessas}
Leach P G L, Cotsakis S and Flessas G P, Symmetry, Singularities and
Integrability in Complex Dynamics I: The Reduction Problem,
{\it J. Nonlin. Math. Phys.}, 2000, V.7, N~4, 445--479.
\bibitem{II}
Leach P G L, Cotsakis S and Flessas G P, Symmetry, Singularities
and Integrabily in Complex Dynamical Systems II: Rescaling and
Time-Translation in Two-Dimensional Systems, {\it J.~Math. Anal.
Appl.} 2000, V.251, 587--608.
\bibitem{liedgl}
Lie S, Differentialgleichungen, Teubner, Leipzig, 1891 (reprinted
Chelsea, New York, 1967).
\bibitem{lieber}
Lie S, Ber\"uhungstransformationen, Teubner, Leipzig, 1896
(reprinted Chelsea, New York, 1977).
\bibitem{liouville}
Liouville J J, Sur l'int\'egration des \'equations diff\'erentielles de la
Dynamique, {\it J. Math. Pures Appl.}, 1855, V.20, 137--138.
\bibitem{leach1}
Mahomed F M and Leach P G L, Symmetry Lie Algebras of $n$th Order
Ordinary Differential Equations, {\it J.
Math. Anal. Appl.}, 1990, V.151,  80--107.
\bibitem{Matzner1}
Matzner R A, 3-Sphere ``Backgrounds'' for the Space Sections of the
Taub Cosmological Solution, {\it J. Math. Phys.}, 1968, V.9, 1063--1066.
\bibitem{leach3}
Mellin C M, Mahomed F M and Leach P G L, Solutions of Generalised
Emden--Fowler
Equations with Two Symmetries, {\it Int. J. Nonlin. Mech.}, 1994, V.29, 529--538.
\bibitem{noether}
Noether E, Invariante Variationsprobleme, {\it Koenig. Gess. Wissen. Nach.
Math.-Phys. K}, 1918, Heft~2, 235--269.
\bibitem{pinney}
Pinney E, The Nonlinear Differential Equation $y''(x)+p(x)y+cy^{-3}=0$, {\it Proc.
Amer. Math. Soc.}, 1950, V.1, 681.
\bibitem{sc81}
Sarlet W and Cantrijn F, Generalizations of Noether's Theorem in
Classical Mechanics, {\it SIAM Review}, 1981, V.23, 467--494.
\bibitem{Spero1}
Spero A and Baierlein R,  Approximate Symmetry Groups of
Inhomogeneous Metrics, {\it J. Math. Phys.}, 1977, V.18, 1330--1340.
\bibitem{Spero2}
Spero A and Baierlein R,  Approximate Symmetry Groups of
Inhomogeneous Metrics: Examples, {\it J. Math. Phys.}, 1978, V.19, 1324--1334.
\bibitem{Zalaletdinov}
Zalaletdinov R, Approximate Symmetries in General Relatvity,
Preprint, International Center for Relativistic Astrophysics,
Dipartamento di Fisica, Universit\'a di Roma ``La Sapienza'',
P~Aldo Moro 5, Roma 00185, Italy.
\end{thebibliography}
\end{document}